\documentclass[prl,twocolumn,superscriptaddress]{revtex4-2}
\usepackage{amsmath}
\usepackage{amssymb}
\usepackage{color}
\usepackage{graphicx}% Include figure files
\usepackage[colorlinks,bookmarks=false,citecolor=blue,linkcolor=red,urlcolor=blue]{hyperref}

\newcounter{fnnumber}

\newcommand{\dd}{\mathrm{d}}

\newcommand{\mbb}[1]{\mathbb{#1}}

\newcommand{\mr}[1]{\mathrm{#1}}
\newcommand{\pd}{\partial}

\begin{document}
\title{Universal behavior in entanglement entropy reveals quantum criticality and underlying symmetry breaking}

\author{Zhe Wang}
\thanks{These authors contributed equally to this work.}
\affiliation{Department of Physics, School of Science and Research Center for Industries of the Future, Westlake University, Hangzhou 310030, China}
\affiliation{Institute of Natural Sciences, Westlake Institute for Advanced Study, Hangzhou 310024, China}

\author{Zehui Deng}
\thanks{These authors contributed equally to this work.}
\affiliation{Beijing Computational Science Research Center, Beijing 100193, China}

\author{Zenan Liu}
\affiliation{Department of Physics, School of Science and Research Center for Industries of the Future, Westlake University, Hangzhou 310030, China}
\affiliation{Institute of Natural Sciences, Westlake Institute for Advanced Study, Hangzhou 310024, China}

\author{Zhiyan Wang}
\affiliation{Department of Physics, School of Science and Research Center for Industries of the Future, Westlake University, Hangzhou 310030, China}
\affiliation{Institute of Natural Sciences, Westlake Institute for Advanced Study, Hangzhou 310024, China}

\author{Yi-Ming Ding}
\affiliation{Department of Physics, School of Science and Research Center for Industries of the Future, Westlake University, Hangzhou 310030, China}
\affiliation{Institute of Natural Sciences, Westlake Institute for Advanced Study, Hangzhou 310024, China}

\author{Long Zhang}
\email{longzhang@ucas.ac.cn}
\affiliation{Kavli Institute for Theoretical Sciences and CAS Center for Excellence in Topological Quantum Computation,
University of Chinese Academy of Sciences, Beijing 100190, China}

\author{Wenan Guo}
\email{waguo@bnu.edu.cn}
\affiliation{School of Physics and Astronomy, Beijing Normal University, Beijing 100875, China}
\affiliation{Key Laboratory of Multiscale Spin Physics (Ministry of Education), Beijing Normal University, Beijing 100875, China}

\author{Zheng Yan}
\email{zhengyan@westlake.edu.cn}
\affiliation{Department of Physics, School of Science and Research Center for Industries of the Future, Westlake University, Hangzhou 310030, China}
\affiliation{Institute of Natural Sciences, Westlake Institute for Advanced Study, Hangzhou 310024, China}

\begin{abstract}
Entanglement takes a key role in quantum physics while how much information it can extract from many-body systems is still an open question, particularly for quantum criticalities and emergent symmetries.
In this work, we systematically study the entanglement entropy (EE) and its derivative (DEE) near quantum phase transitions in various quantum many-body systems. An  one-parameter scaling relation between the DEE and system size at the critical point has been derived for the first time, which successfully obtains the critical exponent via data collapse. What's more, we find that the EE peaks at the (emergent) symmetry-enhanced first-order transition which reflects higher symmetry breaking. This work provides a new paradigm for quantum many-body research from the perspective of EE and DEE.
%Therefore, the EE proves to be a useful information-theoretic measure of quantum critical phenomena. 
%While the quantum critical singular contribution to the EE is of the same order as the nonsingular part, we observe a stronger singularity in the DEE and derive a universal one-parameter scaling relation, which is corroborated by our numerical results. We also find that the EE peaks at the symmetry-enhanced first-order transition which reflects higher symmetry breaking, although the scaling relation fails. Therefore, the EE proves to be a useful information-theoretic measure of quantum critical phenomena.
\end{abstract}

\date{\today}
\maketitle
%\textit{\color{blue} Significance Statement.---} 
%Entanglement entropy (EE) and its derivative (DEE) serve as powerful tools for probing quantum criticality in two-dimensional quantum many-body systems. We demonstrate that DEE exhibits a universal scaling behavior near (2+1)D quantum critical points, revealing stronger singularities than EE itself, and validate this through numerical simulations across O(N) universality classes. While the scaling fails at first-order transitions, EE peaks at symmetry-enhanced transitions due to emergent Goldstone modes, linking entanglement to symmetry-breaking mechanisms. This work establishes EE and DEE as foundational information-theoretic measures for quantum phase transitions, advancing the characterization of critical phenomena beyond conventional order parameters.

\textit{\color{blue} Introduction.---} The information-theoretic perspective of quantum many-body systems is a blooming interdisciplinary research frontier~\cite{Amico2008, Eisert2010, Laflorencie2016, Zeng2019}, where the entanglement entropy (EE) plays a pivotal role. For a subsystem embedded in a large system, the EE quantifies the quantum correlations of the subsystem with the rest of the whole system and provides a generic upper bound of local operator correlations~\cite{Wolf2008}. It grows with the subsystem size following the area law at the ground state of a generic short-range correlated system with an energy gap~\cite{Hastings2007a, Hastings2007, Wolf2008}, and has a logarithmic enhancement beyond the area law in one-dimensional (1D) quantum critical states described by conformal field theory (CFT)~\cite{Holzhey1994, Vidal2003, Korepin2004, Calabrese2004} and gapless fermion systems with Fermi surfaces in higher dimensions~\cite{Wolf2006, Gioev2006a, Ding2012,jiang2024high}. The area law of the EE is the guiding principle for the development of numerical schemes based on matrix-product and tensor network states~\cite{Verstraete2004, Verstraete2008a, Schollwock2011a, Cirac2021}. Moreover, the strong subadditivity of the EE was adopted to prove the monotonicity of the renormalization group (RG) flow in relativistic quantum field theory~\cite{Casini2007, Casini2012}.

%The area law of the EE arises from the short-range correlations in the gapped phase  both the short-range correlations across the boundary and the long-range correlations, with the latter becoming particularly significant and singular close to a quantum critical point (QCP) with a diverging correlation length. 

In a 2D quantum many-body system with (emergent) Lorentz symmetry, the area law of the EE states that $S(\ell)\sim \alpha \ell$, in which $\ell$ is the length of the subsystem boundary. The area law of the EE arises from short-range correlations across the boundary in the gapped disordered phase. In addition, it also arises from long-range correlations near a quantum critical point (QCP) where the correlation length diverges. The prefactor $\alpha$ varies with the tuning parameter $g$ in the Hamiltonian and exhibits a cusp singularity at the QCP $g_{c}$~\cite{Metlitski2009}, $\alpha(g)-\alpha(g_{c})\propto |g-g_{c}|^{\nu}$, in which $\nu$ is the correlation length critical exponent. Such a cusp singularity has been confirmed through quantum Monte Carlo (QMC) simulations and mean-field theory for (2+1)D QCPs~\cite{Helmes2014, Frerot2016}. However, a more complete quantum critical scaling form of the EE itself (not only the prefactor $\alpha$)  and its derivative remains largely elusive, due to the lack of highly efficient numerical methods for computing the EE in 2D quantum many-body systems.

In this work, we investigate the universal scaling behavior of the EE and its derivative (denoted by DEE) in the vicinity of (2+1)D QCPs. The EE is calculated with the bipartite reweight-annealing QMC scheme developed recently~\cite{Wang2024b,ding2024tracking,wang2024addressing} by applying a reweight-annealing procedure~\cite{Ding2024, Ma2024} along the path of real physical parameters, which enables the high-throughput scan of the EE near the QCP. While the cusp singularity of the EE is at the same order as its nonsingular part, we observe a stronger singularity in the DEE data near the QCP. We propose a universal one-parameter scaling relation for the DEE in Eq.~(\ref{eq:dell}), which is corroborated by the data collapse near (2+1)D QCPs in the O($N$) ($1\leq N\leq 3$) universality classes. The scaling relation fails for first-order quantum phase transitions; Nonetheless, we find the EE shows a peak at the first-order transition that spontaneously breaks an enhanced continuous symmetry, which can be attributed to the larger number of Goldstone modes at the transition. Therefore, our work demonstrates that the EE and its derivative are useful information-theoretic measures of quantum phase transitions.

\begin{figure}[!tb]
\centering
\includegraphics[width=\columnwidth]{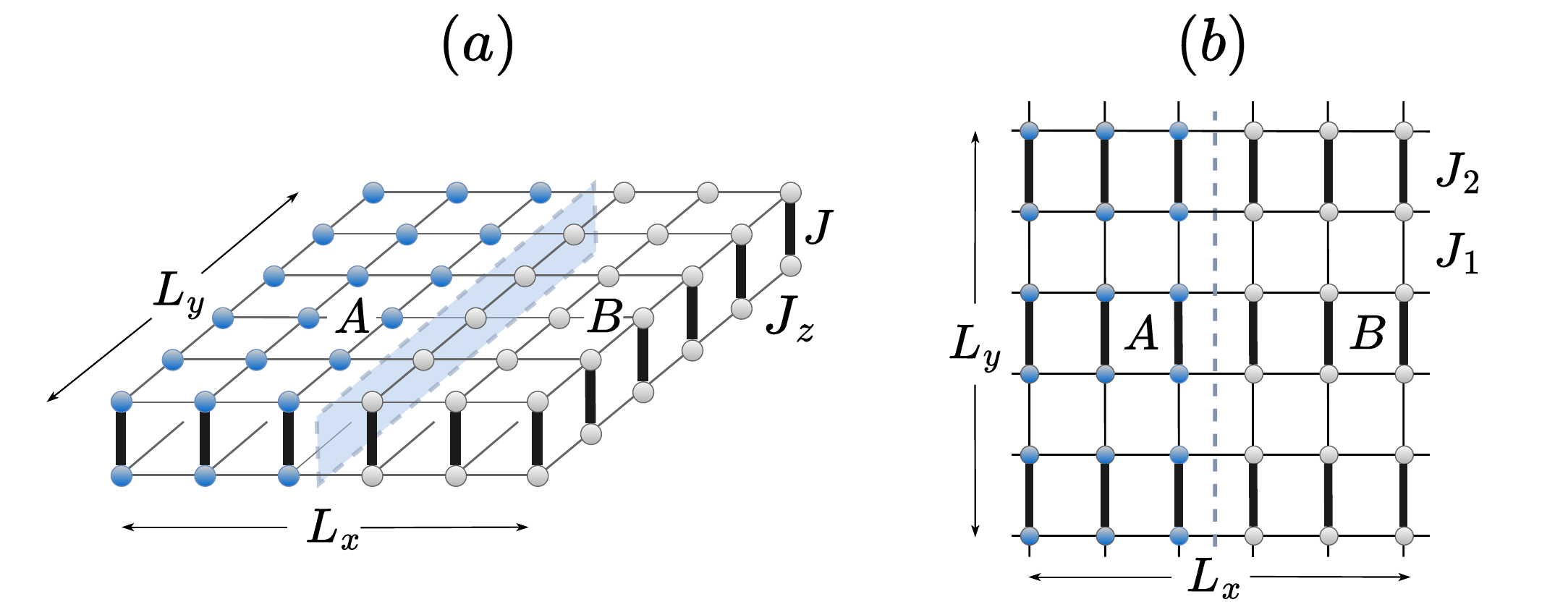}
\caption{Quantum spin models on a square lattice with periodic boundary condition. The lattice size is $L_{x}\times L_{y}$, and the subsystem is a $(L_{x}/2)\times L_{y}$ cylinder region, which is separated from the rest part by the dashed lines. (a) Spin-1/2 bilayer Ising-Heisenberg model. $-J_{z}< 0$ denotes the intralayer FM Ising interaction, while $J>0$ is the interlayer AF Heisenberg interaction. (b) Spin-1/2 dimerized Heisenberg model. The AF interaction strengths on the thin and the thick bonds are $J_{1}$ and $J_{2}$, respectively. The anisotropy parameter $\Delta$ is the same on all bonds.}
\label{fig:model}
\end{figure}

\textit{\color{blue} Models and method.---} We will focus on quantum spin models with QCPs in the (2+1)D O($N$) ($1\leq N\leq 3$) universality classes. We also study the symmetry-enhanced first-order transitions for comparison. The model Hamiltonians are illustrated in Fig.~\ref{fig:model} and described in detail in the following sections. Numerical simulations are performed on $L_{x}\times L_{y}$ square lattice with periodic boundary conditions in both directions. Considering the dynamical critical exponent $z=1$ in this work, the inverse temperature increases linearly with the lattice size, $\beta=L_{x}$, in order to approach the quantum critical regime. The subsystem is the $(L_{x}/2)\times L_{y}$ cylinder region with two smooth boundaries, and the total boundary length $\ell=2L_{y}$. We adopt the bipartite reweight-annealing algorithm~\cite{Wang2024b} combined with the stochastic series expansion (SSE) QMC simulations~\cite{Sandvik1992a, Sandvik1999, Syljuasen2002, Sandvik2010a, Yan2019a, Yan2022} to extract the second R\'enyi EE of the subsystem, $S^{(2)}=-\ln\mr{tr}\rho_{A}^{2}$, in which $\rho_{A}$ is the reduced density matrix. The calculation details are presented in the Supplemental Materials (SM)~\footnote{See the Supplemental Materials for further details of the reweight-annealing algorithm for the R\'enyi EE, the definition of the checkerboard $J$-$Q$ model, and our failed attempts at data collapse analysis of the EE near quantum critical points and the derivative of EE near first-order transitions, and Refs.~\cite{Huang2024a}.}\setcounter{fnnumber}{\thefootnote}.

\textit{\color{blue} Universal scaling relation.---} Let us first derive a universal quantum critical scaling form for the EE and its derivative, which is valid for the R\'enyi EE as well as for the von Neumann EE. In the quantum critical regime, the EE is contributed from both the nonsingular short-range correlations and the singular long-range correlations. For a subsystem with a smooth boundary, the EE near the QCP is given by~\cite{Metlitski2009}
\begin{equation}
S(\ell, g)= a(g)\ell +\tilde{S}_{0}(x),
\label{eq:sell}
\end{equation}
in which $\ell$ is the boundary length. The prefactor $a(g)$ in the first term varies analytically with the tuning parameter $g$, while $\tilde{S}_{0}$ is a universal function of the dimensionless scaling variable $x=(g-g_{c})\ell^{1/\nu}$~\footnote{The universal scaling function in the EE was denoted by $S_{0}(\ell/\xi)$ as a function of the dimensionless ratio $\ell/\xi$ in Ref.~\cite{Metlitski2009}, in which $\xi\propto |g-g_{c}|^{-\nu}$ is the correlation length. Here, we introduce the scaling variable $x=(g-g_{c})\ell^{1/\nu}\propto (\ell/\xi)^{1/\nu}$ to facilitate further analysis.}. At the QCP, the $\tilde{S}_{0}$ term reduces to a universal constant, which only depends on the universality class of the QCP and the geometric invariants of the subsystem, e.g., the aspect ratio of the cylinder region. Away from the QCP, the $\tilde{S}_{0}$ term also contributes to the area law, $\tilde{S}_{0}(x)=r|g-g_{c}|^{\nu}\ell$ for $|x|\gg 1$, in which $r$ is a constant coefficient, hence leading to a cusp singularity of the area-law prefactor, $\alpha(g)=a(g)+r|g-g_{c}|^{\nu}$~\cite{Metlitski2009}. Therefore, both terms in Eq.~(\ref{eq:sell}) are of the same order in the $\ell\rightarrow\infty$ limit and are difficult to be clearly separated in numerical analysis.

\begin{figure*}[!tb]
\centering
\includegraphics[width=\textwidth]{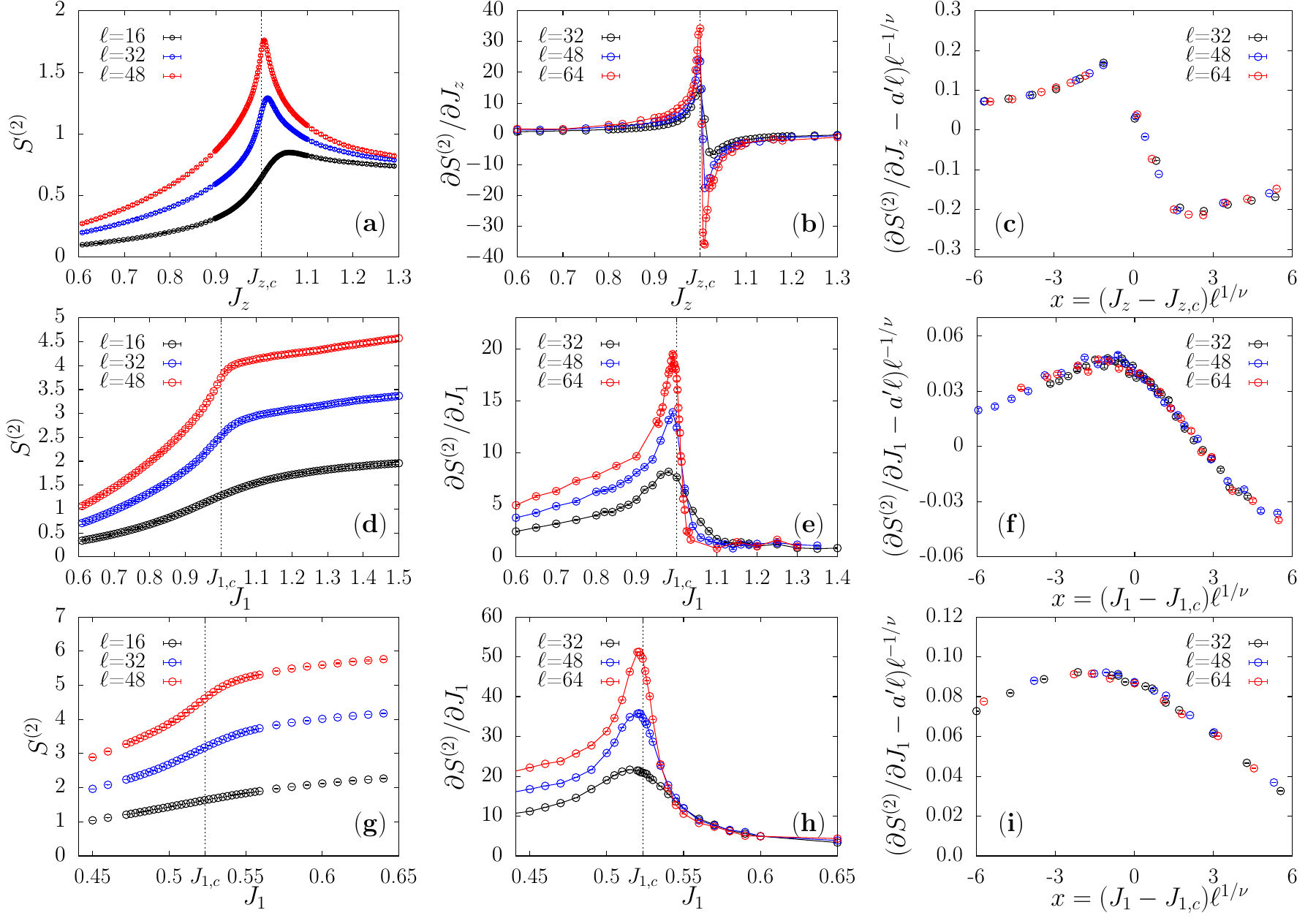}
\caption{(a) The second R\'enyi EE and (b) its derivative, and (c) the scaling function $\tilde{S}_{0}'(x)$ obtained from the data collapse analysis of the DEE near the Ising transition of the bilayer Ising-Heisenberg model. In the calculations, $J=3.045$ is fixed, while $J_{z}$ is tuned near the QCP $J_{z,c}=1$. (d--f) Data of the dimerized Heisenberg model with the anisotropy parameter $\Delta=0.9$. $J_{2}=2.1035$ is fixed, while $J_{1}$ is tuned around the O(2) QCP $J_{1,c}=1$. (g--i) Data of the dimerized isotropic Heisenberg model. $J_{2}=1$ is fixed, while $J_{1}$ is tuned around the O(3) QCP $J_{1,c}=0.52337$.}
\label{fig:qcp}
\end{figure*}

From our numerical results shown in Fig.~\ref{fig:qcp}, we observe that the derivative of the EE over the tuning parameter $\pd S(\ell,g)/\pd g$ exhibits a stronger singularity near the QCP. From Eq.~(\ref{eq:sell}), we find
\begin{equation}
\frac{\pd S(\ell,g)}{\pd g}= a'(g)\ell + \tilde{S}'_{0}(x)\ell^{1/\nu}.
\label{eq:dell}
\end{equation}
For the (2+1)D O($N$) ($1\leq N\leq 3$) models studied in this work, the critical exponent $\nu<1$, thus the second term in Eq.~(\ref{eq:dell}) derived from the critical behavior dominates in the $\ell\rightarrow \infty$ limit. Therefore, the DEE is a better diagnosis of the universal scaling behavior of the quantum entanglement in the quantum critical regime. The scaling function $\tilde{S}_{0}'(x)$ can be expanded into a polynomial near the QCP,
\begin{equation}
\tilde{S}_{0}'(x)=\sum_{k=0}^{k_{\mr{max}}}\frac{1}{k!}\tilde{S}_{0}^{(k+1)}(0)x^{k}.
\end{equation}
The first term in Eq.~(\ref{eq:dell}) serves as a subleading correction to the scaling relation of the DEE. The prefactor $a'(g)$ can be expanded into a power series in general,
\begin{equation}
a'(g)=a'(g_{c})+a''(g_{c})(g-g_{c})+\ldots,
\end{equation}
and we find it suffices to only retain the constant term in practice. The critical point $g_{c}$, the critical exponent $\nu$, and the coefficients $\tilde{S}_{0}^{(k+1)}(0)$ and $a'(g_{c})$ are the fitting parameters in the following data collapse analysis.

\textit{\color{blue} Ising transition.---} Let us focus on the spin-1/2 bilayer Ising-Heisenberg model shown in Fig.~\ref{fig:model}~(a). Its Hamiltonian is given by~\cite{Wu2023b}
\begin{equation}
H=-J_{z}\sum_{\alpha=1,2}\sum_{\langle ij\rangle}S_{i,\alpha}^{z}S_{j,\alpha}^{z}+J\sum_{i}\vec{S}_{i,1}\cdot\vec{S}_{i,2},
\end{equation}
in which $\alpha$ is the layer index, and $\langle ij\rangle$ denotes the summation over nearest-neighbor bonds within each layer. $-J_{z}<0$ is the intralayer ferromagnetic (FM) Ising interaction, while $J>0$ is the interlayer antiferromagnetic (AF) Heisenberg interaction. For $J/J_{z}\ll 1$, the ground state has an intralayer-FM-interlayer-AF order, which breaks the $\mbb{Z}_{2}$ spin flip symmetry generated by $X=\prod_{i,\alpha}\sigma_{i,\alpha}^{x}$. For $J/J_{z} \gg 1$, the ground state approaches a direct product of the interlayer spin-singlet states, which has an energy gap and does not break any symmetries. The QCP separating these two phases is located at $J/J_{z}=3.045(2)$ and belongs to the (2+1)D Ising universality class~\cite{Wu2023b}.

Fixing $J=3.045$ and tuning $J_{z}$ around the QCP $J_{z,c}=1$, we evaluate the second R\'enyi EE and its derivative for different lattice sizes. The results are plotted in Figs.~\ref{fig:qcp}~(a) and (b). The EE shows a maximum at the QCP, thus the DEE changes sign across the QCP. The maximum of the EE can be attributed to the stronger spatial correlations in the quantum critical state than in the adjacent gapped phases.

We cannot achieve a satisfactory data collapse for the EE data with the dimensionless scaling variable $x=(J_{z}-J_{z,c})\ell^{1/\nu}$~\footnotemark[\thefnnumber], because the singular and the nonsingular contributions are of the same order as we argued before. Nonetheless, the stronger singularity in the DEE data motivates us to perform scaling analysis on the DEE. In this case, it is difficult to capture the scaling function $\tilde{S}_{0}'(x)$ with a single (low-order) polynomial on both sides of the QCP due to the abrupt sign change of the DEE. Hence, we first fit the scaling relation Eq.~(\ref{eq:dell}) to the DEE data on the right-hand side of the QCP ($J_{z}>1$) and obtain $J_{z,c}=1.003(1)$ and $\nu=0.62(7)$. Meanwhile, we find the DEE data on the left-hand side ($J_{z}<1$) also collapse perfectly onto a single curve of the scaling variable $x$ with the same $J_{z,c}$ and $\nu$ [see Fig.~\ref{fig:qcp}~(c)]. Moreover, these values are fully consistent with the QCP $J_{z,c}=1$ in the literature~\cite{Wu2023b} and the critical exponent $\nu=0.63012$ of the (2+1)D Ising universality class~\cite{Guida1998, Deng2003, Wu2023b}. The scaling function $\tilde{S}'_{0}(x)$ obtained from the data collapse analysis is shown in Fig.~\ref{fig:qcp}~(c).

\textit{\color{blue} Continuous symmetry breaking transitions.---} We then study the spin-1/2 dimerized Heisenberg model illustrated in Fig.~\ref{fig:model}~(b). The Hamiltonian is given by~\cite{Matsumoto2001a, Zhu2021b}
\begin{equation}
H=J_{1}\sum_{\langle ij\rangle}D_{ij}+J_{2}\sum_{\langle ij\rangle'}D_{ij},
\label{eq:hei}
\end{equation}
in which the summation in the $J_{1}$-term is taken over the thin bonds, while the $J_{2}$-term is over the thick bonds. The nearest-neighbor AF Heisenberg interaction $D_{ij}=S_{i}^{x}S_{j}^{x}+S_{i}^{y}S_{j}^{y}+\Delta S_{i}^{z}S_{j}^{z}$ is in general anisotropic, and we consider a spatially uniform anisotropy $\Delta$ in the range $0<\Delta\leq 1$. In the $J_{2}/J_{1}\simeq 1$ regime, the ground state has a long-range AF order. For $0<\Delta<1$, the AF order parameter lies in the $xy$-plane and breaks the O(2) spin rotation symmetry, while for $\Delta=1$, the interaction is isotropic, and the AF order breaks the O(3) symmetry. In the $J_{2}/J_{1}\gg 1$ limit, the ground state approaches a direct product of the spin-singlet states on the thick bonds, which has an energy gap and does not break any symmetries. The QCP separating the gapped phase and the AF ordered phase belongs to the (2+1)D O(2) ($0<\Delta<1$) or the O(3) ($\Delta=1$) universality class.

The second R\'enyi EE and its derivative are evaluated around the QCPs in the easy-plane case ($\Delta=0.9$) and in the isotropic case ($\Delta=1$). The results are plotted in Figs.~\ref{fig:qcp}~(d, e) and (g, h), respectively. In each case, we fix $J_{2}$ and tune $J_{1}$ around the QCP as described in the figure caption. We find the EE grows monotonically from the gapped phase across the QCP to the AF-ordered phase, which is quite different from the Ising transition case. This indicates that the quantum correlations in the AF ordered phase are even stronger than at the QCP, which can be attributed to the gapless Goldstone modes (spin waves) due to the continuous symmetry breaking~\cite{Song2011b, Kallin2011a, Metlitski2011}. Similar behavior was observed near the QCP of the bilayer Heisenberg model in the (2+1)D O(3) universality class~\cite{Helmes2014}.

Similar to the Ising transition case, we cannot achieve a satisfactory data collapse for the EE either~\footnotemark[\thefnnumber], thus we turn to the DEE. Fitting the scaling relation Eq.~(\ref{eq:dell}) to the DEE data, we find $J_{1,c}=0.997(2)$ and $\nu=0.66(2)$ in the easy-plane case ($\Delta=0.9$), and $J_{1,c}=0.523(1)$ and $\nu=0.71(4)$ in the isotropic case ($\Delta=1$). These results are fully consistent with the known QCPs and the critical exponents: $J_{1,c}=1$ and $\nu=0.6703$ for the O(2) transition ($\Delta=0.9$), and $J_{1,c}=0.52337$ and $\nu=0.7073$ for the O(3) transition ($\Delta=1$)~\cite{Guida1998, Matsumoto2001a, Zhu2021b, Zhang2017}. The scaling functions $\tilde{S}_{0}(x)$ obtained from the data collapse analysis are plotted in Figs.~\ref{fig:qcp}~(f) and (i).

\begin{figure}[!tb]
\centering
\includegraphics[width=\columnwidth]{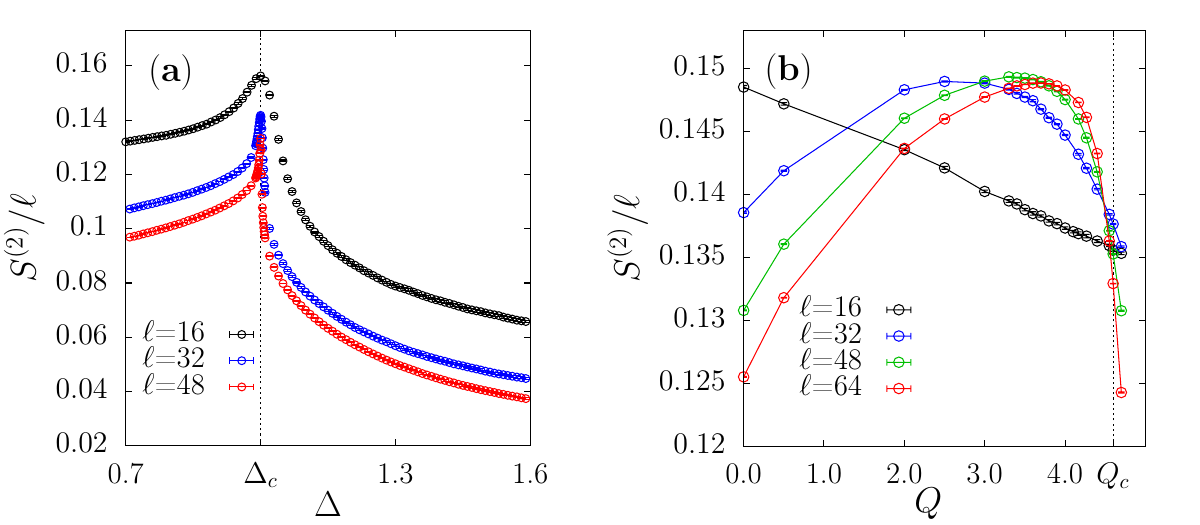}
\caption{The second R\'enyi EE divided by the boundary length $S^{(2)}/\ell$ of (a) the anisotropic Heisenberg model and (b) the checkerboard $J$-$Q$ model near the symmetry-enhanced first-order transitions.}
\label{fig:first-order}
\end{figure}

\textit{\color{blue} Symmetry-enhanced first-order transitions.---} Having established the universal scaling relation of the DEE near (2+1)D QCPs, we examine the first-order quantum phase transitions with enhanced symmetries for comparison.

Let us first study the 2D anisotropic Heisenberg model given in Eq.~(\ref{eq:hei}) with a spatially uniform AF interaction $J_{1}=J_{2}=1$, and take the anisotropy $\Delta$ as tuning parameter. At $\Delta=1$, the Heisenberg interaction is isotropic with an O(3) symmetry, which is spontaneously broken by the AF order at the ground state. Away from this point, the symmetry is explicitly reduced to $\mr{O}(2)=\mr{SO}(2)\rtimes \mbb{Z}_{2}$ generated by the uniaxial spin rotation and the spin-flip transformations. For $0<\Delta<1$, the ground state has an easy-plane AF order and breaks the SO(2) spin rotation symmetry, while for $\Delta>1$, the ground state has an easy-axis AF order and breaks the $\mbb{Z}_{2}$ spin flip symmetry. The easy-plane and the easy-axis AF order parameters coexist at $\Delta=1$, and jump to zero across the transition, hence this is a first-order transition with a spontaneously broken enhanced symmetry~\cite{Zhao2019}.

The second R\'enyi EE near the transition is plotted in Fig.~\ref{fig:first-order}~(a). It shows a sharp peak at the transition point indicating stronger quantum correlations than the adjacent phases, which can be attributed to the larger number of Goldstone modes and low-energy density of states (DOS) at the transition point. We cannot achieve a satisfactory data collapse for the DEE data~\footnotemark[\thefnnumber], which is not surprising given the first-order nature of the transition.

The symmetry-enhanced first-order transition is not restricted to systems with exact symmetries. An enhanced symmetry can emerge in the long-wavelength limit if all explicit symmetry-breaking interactions are irrelevant and diminish with the RG flow. Such an enhanced symmetry can even emerge at a first-order transition between two ordered phases. At the putative deconfined QCP between the AF order and the valence-bond solid (VBS) order~\cite{Senthil2004b, Senthil2004a} in the $J$-$Q$ model~\cite{Sandvik2007, Lou2009a}, an emergent SO(5) symmetry of the AF and the VBS order parameters was observed~\cite{Nahum2015a, Takahashi2020, Takahashi2024}. While theoretical arguments~\cite{Senthil2004b, Senthil2004a, Senthil2006a} and early numerical results~\cite{Sandvik2007, Lou2009a} suggested a continuous quantum phase transition, anomalous scaling behavior was observed in various physical quantities, and its origin has been debated for a long time~\cite{Kuklov2008, Jiang2008b, Sandvik2010b, Chen2013f, Shao2016a, Zhao2020a, Sandvik2020a, DEmidio2023}. Recently, the anomalous logarithmic correction term in the EE at the transition~\cite{Zhao2022a, Song2023, Liao2023, DEmidio2024, Song2024a} has been observed, and it is attributed to Goldstone modes associated with continuous symmetry breaking. The number of Goldstone modes is consistent with the spontaneous breaking of an (emergent) SO(5) symmetry. Hence, it supports the symmetry-enhanced first-order transition scenario~\cite{Deng2024}.

%Recently, the anomalous logarithmic correction term in the EE at the transition~\cite{Zhao2022a, Song2023, Liao2023, DEmidio2024, Song2024a} was found consistent with the spontaneous breaking of an (emergent) SO(5) symmetry, hence it corroborates the symmetry-enhanced first-order transition scenario~\cite{Deng2024}.

We study the checkerboard $J$-$Q$ model on a square lattice, which has a first-order transition between the O(3)-symmetry-breaking AF order and the $\mbb{Z}_{2}$-symmetry-breaking plaquette-singlet solid (PSS) order at $Q_{c}/J=4.598(1)$~\cite{Zhao2019}~\footnotemark[\thefnnumber]. At the transition point, an enhanced O(4) symmetry among the AF and the PSS order parameters emerges and is spontaneously broken at the ground state~\cite{Zhao2019}. Its R\'enyi EE near the transition point is plotted in Fig.~\ref{fig:first-order}~(b). The EE shows a broad peak, which progressively moves to the transition point as the lattice size increases. This is qualitatively consistent with the spontaneous breaking of the enhanced symmetry with more Goldstone modes and low-energy DOS at the transition than in the adjacent AF and PSS phases. The drift of the peak implies that the enhanced symmetry is not exact but emergent with the RG flow. Specifically, there exists a relatively large irrelevant field, which typically scales as a negative power of the system size. As the system size increases, the effect of this irrelevant field gradually vanishes. Consequently, we observe that the peak position corresponding to the enhanced symmetry shifts  with increasing system size. This is also reflected in the fact that, as the system size increases, the width of the peak gradually narrows. In the thermodynamic limit, it should evolve into a sharp peak.% Similar to the case of the anisotropic Heisenberg model, we do not find a satisfactory data collapse for the DEE data either.

The relationship of the EE and the low-energy DOS can be explained further based on the correspondence between the entanglement spectrum (ES) and the low-energy spectrum of the subsystem with a boundary. Such a correspondence was first demonstrated~\cite{Li2008, Qi2012} for topological states with gapless boundary, and was also rigorously proved in (1+1)D CFT~\cite{Lauchli2013, Cardy2016, Yu2022}. Recently, such a correspondence was put forward in a more general setting based on the ``wormhole picture'' in the path integral formulation of the reduced density matrix and verified with numerical simulations~\cite{Yan2023b, Song2023a, Liu2024b}. Thereby, a larger low-energy DOS from the Goldstone modes implies a denser low-energy ES and hence a larger EE.

\textit{\color{blue} Conclusion and discussions.---} We have systematically studied the EE and its derivative around quantum phase transitions in 2D quantum many-body systems. In the quantum critical regime, although the singular and the nonsingular terms in the EE are of the same order and cannot be clearly separated, the DEE shows a stronger singularity induced by the quantum criticality. We have obtained a universal scaling relation for the DEE, and corroborated it with the data collapse near (2+1)D QCPs. The scaling relation fails for first-order transitions; Nonetheless, the EE shows a peak at symmetry-enhanced first-order transitions, which can be attributed to the larger low-energy DOS from the Goldstone modes.

The EE is a basis-independent measure of generic quantum correlations, hence it does not require a prior knowledge of the quantum phase transitions and should be particularly suitable for detecting exotic QCPs beyond the conventional paradigm of spontaneous symmetry breaking. Other information-theoretic measures of quantum many-body systems may also be explored to characterize quantum critical phenomena in the future.

\begin{acknowledgements}
\textit{\color{blue} Acknowledgements.---} We thank the helpful discussions with Marchello Delmonte, Yan-Cheng Wang, Yin Tang, and Wei Zhu. W.G. thanks the support from the National Natural Science Foundation of China under Grant No.~12175015. L.Z. is supported by the National Natural Science Foundation of China (No.~12174387), the Chinese Academy of Sciences (Nos.~YSBR-057 and JZHKYPT-2021-08), and the Innovative Program for Quantum Science and Technology (No.~2021ZD0302600).  Z.W. and Z.Y. are supported by the start-up funding of Westlake University and the China Postdoctoral Science Foundation under Grant No.~2024M752898. The authors thank the high-performance computing center of Westlake University and the Beijing PARATERA Tech Co., Ltd. for providing HPC resources.
\end{acknowledgements}

\providecommand{\newblock}{}

\clearpage
\appendix
\setcounter{equation}{0}
\setcounter{figure}{0}
\renewcommand{\theequation}{S\arabic{equation}}
\renewcommand{\thefigure}{S\arabic{figure}}
\setcounter{page}{1}
\begin{widetext}
%\linespread{1.05}
	
\centerline{\bf\Large Supplemental Materials}

\section{Bipartite reweight-annealing algorithm}

The $n$-th R\'enyi entanglement entropy (EE) of a subsystem $A$ coupled with an environment $B$ is defined by $S^{(n)}=\frac{1}{1-n}\ln\big(\mr{Tr}(\rho_{A}^{n})\big)$, in which $\rho_{A}=\mr{Tr}_{B}\rho$ is the reduced density matrix. If the whole system is in a thermal equilibrium state, $\rho=e^{-\beta H}/Z$, the R\'enyi EE can be obtained by $\mr{Tr}_{A}\rho_{A}^{(n)}=Z_{A}^{(n)}/Z^{n}$, in which $Z=\mr{Tr}e^{-\beta H}$ and $Z_{A}^{(n)}=\mr{Tr}_{A}\big((\mr{Tr}_{B}e^{-\beta H})^{n}\big)$ are the partition function and the $n$-th replica partition function, respectively.

The R\'enyi EE is evaluated with the bipartite reweight-annealing algorithm~\cite{Wang2024b} combined with the stochastic series expansion (SSE) quantum Monte Carlo (QMC) simulations~\cite{Sandvik1992a, Sandvik1999, Syljuasen2002, Sandvik2010a, Yan2019a, Yan2022}. In this algorithm, the two partition functions $Z_{A}^{(n)}$ and $Z=Z_{A}^{(1)}$ are computed separately via the reweight-annealing scheme~\cite{Ding2024, Ma2024} along a path of physical parameters $J_{i}$ ($0\leq i\leq N$)~\cite{Ding2024, Ma2024},
\begin{equation}
\frac{Z_{A}^{(n)}(J_{N})}{Z_{A}^{(n)}(J_{0})}=\prod_{i=0}^{N-1}\frac{Z_{A}^{(n)}(J_{i+1})}{Z_{A}^{(n)}(J_{i})}.
\end{equation}
The ratio of two adjacent partition functions is evaluated with the reweighting trick,
\begin{equation}
\frac{Z_{A}^{(n)}(J_{i+1})}{Z_{A}^{(n)}(J_{i})}=\bigg\langle \frac{W(J_{i+1})}{W(J_{i})}\bigg\rangle_{Z_{A}^{(n)}(J_{i})},
\end{equation}
in which $\langle\cdot\rangle_{Z_{A}^{(n)}(J_{i})}$ indicates that the Monte Carlo (MC) sampling and averaging are taken over the replica manifold $Z_{A}^{(n)}$ with the physical parameter $J_{i}$, while $W(J_{i+1})$ and $W(J_{i})$ denote the weights of the same configuration for different parameters $J_{i+1}$ and $J_{i}$, respectively. The reference point $J_{0}$ in the path is chosen such that the ratio $Z_{A}^{(n)}(J_{0})/Z(J_{0})^{n}$ is already known. For example, we may take the Hamiltonian composed of two disconnected subsystems $A$ and $B$ as the reference point, then the whole system forms a direct-product state, $\rho=\rho_{A}\otimes\rho_{B}$, hence the ratio $Z_{A}^{(n)}(J_{0})/Z(J_{0})^{n}=1$. Therefore, this reweight-annealing scheme enables the efficient high-throughput evaluation of the R\'enyi EE along a given path of physical parameters.

The derivative of the $n$-th R\'enyi EE over the physical parameter $J$ is given by~\cite{Wang2024b}
\begin{equation}
\frac{\dd S^{(n)}}{\dd J}=\frac{\beta}{1-n}\bigg(n\Big\langle \frac{\dd H}{\dd J} \Big\rangle_{Z}-\Big\langle \frac{\dd H}{\dd J} \Big\rangle_{Z_{A}^{(n)}} \bigg),
\end{equation}
in which $\langle\cdot\rangle_{Z}$ and $\langle\cdot\rangle_{Z_{A}^{(n)}}$ indicate that the MC sampling and averaging are taken in the configuration spaces for the partition function $Z$ and the replica partition function $Z_{A}^{(n)}$, respectively.

\begin{figure}[!tb]
\centering
\includegraphics[width=0.5\textwidth]{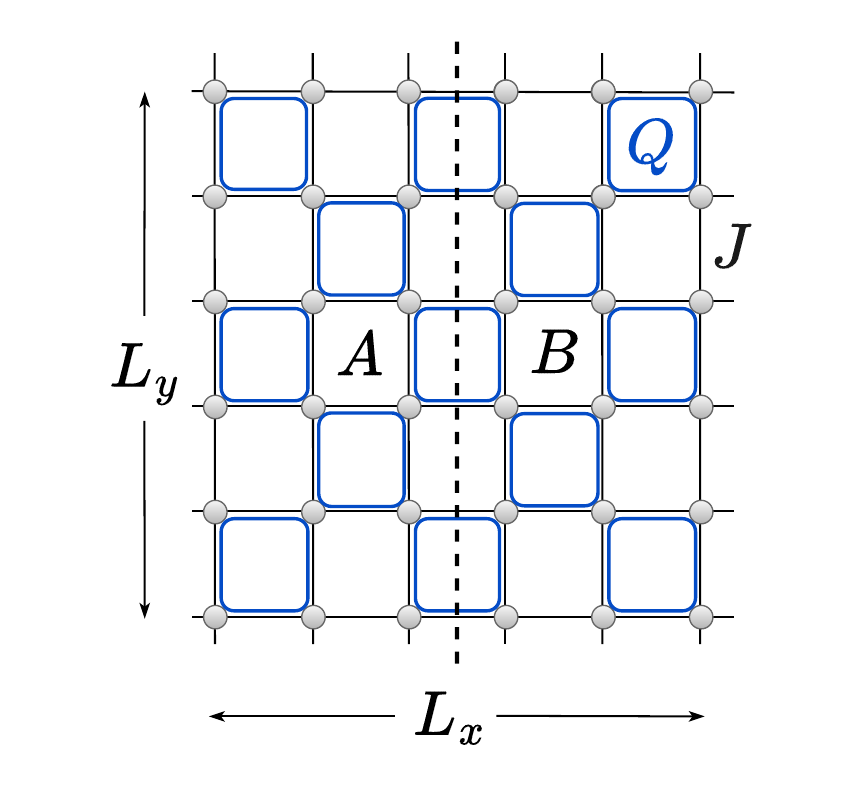}
\caption{The checkerboard $J$-$Q$ model on a $L_{x}\times L_{y}$ square lattice with periodic boundary condition in both directions. The subsystem $A$ is a $(L_{x}/2)\times L_{y}$ cylinder region with the boundary length $\ell=2L_{y}$. The AF Heisenberg interaction ($J$-term) acts on all n.n. bonds, while the four-spin interaction ($Q$-term) acts on n.n. bond pairs that form a shaded plaquette.}
\label{fig:cbjq}
\end{figure}

\section{Checkerboard $J$-$Q$ model}

The Hamiltonian of the checkerboard $J$-$Q$ model on a square lattice is defined by~\cite{Zhao2019}
\begin{equation}
H=-J\sum_{\langle ij \rangle}P_{ij}-Q\sum_{[ijkl]}P_{ij}P_{kl},
\label{eq:cbjq}
\end{equation}
in which $P_{ij}=1/4-\vec{S}_{i}\cdot\vec{S}_{j}$ is the projection operator into the spin singlet state of two sites $i$ and $j$. The first summation in Eq.~(\ref{eq:cbjq}) is taken over all nearest-neighbor (n.n.) bonds on the square lattice, while the second is over the n.n. bond pairs $(ij)$ and $(kl)$ that form a shaded plaquette shown in Fig.~\ref{fig:cbjq}.

The quantum phase diagram was obtained using QMC simulations~\cite{Zhao2019}. For $Q/J\ll 1$, the ground state has an antiferromagnetic (AF) order breaking the O(3) spin rotation symmetry, while for $Q/J\gg 1$, the system forms a two-fold degenerate plaquette-singlet solid (PSS) order, which spontaneously breaks the $\mbb{Z}_{2}$ lattice symmetry. At the first-order transition point $Q_{c}/J=4.598(1)$, the system has an emergent O(4) symmetry among the AF and the PSS order parameters, which is also spontaneously broken by the coexisting AF and PSS orders.

\begin{figure}[!tb]
\centering
\includegraphics[width=\textwidth]{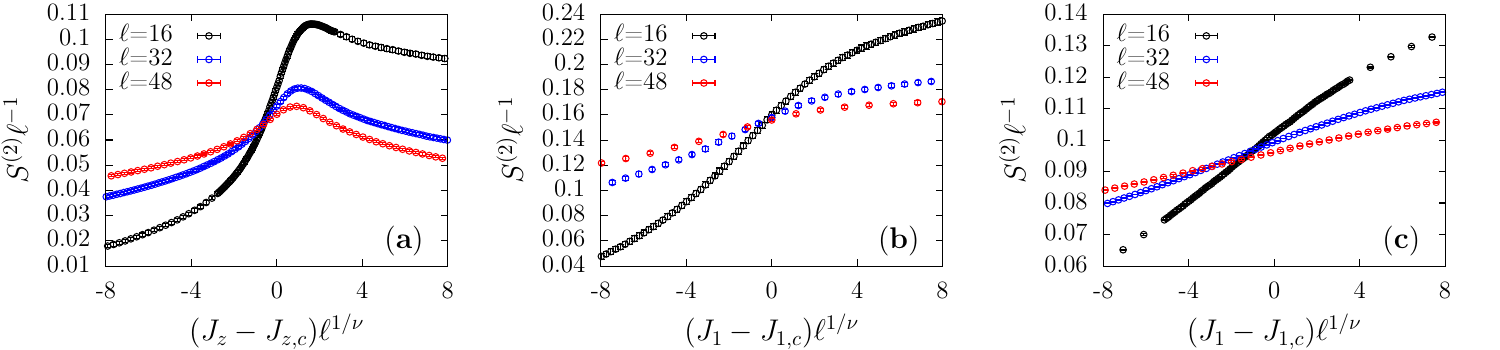}
\caption{Failed data collapse analysis of the second R\'enyi EE data near the (2+1)D O($N$) ($1\leq N\leq 3$) QCPs based on the proposed scaling form Eq.~(\ref{eq:ee})~\cite{Huang2024a}. The quantum spin models with these QCPs are described in the main text. The vertical axis is the rescaled EE $S^{(2)}(\ell,g)\ell^{-1}$, while the horizontal axis is the dimensionless scaling variable $x=(g-g_{c})\ell^{1/\nu}$ calculated with the QCPs and the critical exponents well-established in the literature~\cite{Wu2023b, Matsumoto2001a, Zhu2021b, Guida1998, Deng2003, Zhang2017}: (a) $J_{z,c}=1$ and $\nu=0.630$ for the Ising transition, (b) $J_{1,c}=1$ and $\nu=0.671$ for the O(2) transition, and (c) $J_{1,c}=0.52337$ and $\nu=0.707$ for the O(3) transition. Were the scaling form Eq.~(\ref{eq:ee}) valid, the rescaled data for different lattice sizes would collapse onto a single curve, which is unfortunately not the case.}
\label{fig:ee}
\end{figure}

\section{Failed attempt at scaling analysis of EE data near quantum critical points}

For a two-dimensional (2D) quantum many-body system with (emergent) Lorentz symmetry, the EE of a subsystem with smooth boundary obeys the area law at the ground state, $S(\ell)\sim \alpha\ell$, in which $\ell$ is the boundary length. The prefactor $\alpha$ varies with the tuning parameter $g$ in the Hamiltonian.

In order to unveil the universal critical behavior of the EE near a quantum critical point (QCP), a one-parameter scaling form was proposed in Ref.~\cite{Huang2024a},
\begin{equation}
S(\ell,g)=\ell f(x),
\label{eq:ee}
\end{equation}
which assumes that the critical behavior manifests itself in the area-law prefactor $f(x)$ as a universal function of the dimensionless scaling variable $x=(g-g_{c})\ell^{1/\nu}$. However, we find that this proposed scaling form cannot capture our EE data near the (2+1)D O($N$) ($1\leq N\leq 3$) QCPs, as described in detail in the caption of Fig.~\ref{fig:ee}.

In the quantum critical regime, the EE is contributed from both the nonsingular short-range correlations and the singular long-range correlations. Therefore, the EE of a subsystem with smooth boundary is given by~\cite{Metlitski2009}
\begin{equation}
S(\ell,g)=a(g)\ell+\tilde{S}_{0}(x),
\label{eq:sell}
\end{equation}
in which $a(g)$ varies analytically with the tuning parameter $g$, and $\tilde{S}_{0}$ is a universal function of the dimensionless scaling variable $x=(g-g_{c})\ell^{1/\nu}$ near the QCP. At the QCP, the $\tilde{S}_{0}$ term reduces to a universal constant, which only depends on the universality class of the QCP and the geometric invariants of the subsystem, e.g., the aspect ratio of the cylinder region. Away from the QCP, the $\tilde{S}_{0}$ term also contributes to the area law, $\tilde{S}_{0}(x)=r|g-g_{c}|^{\nu}\ell$ for $|x|\gg 1$, in which $r$ is a constant coefficient, hence leading to a cusp singularity in the prefactor of the area law $\alpha(g)=a(g)+r|g-g_{c}|^{\nu}$~\cite{Metlitski2009}. Therefore, both terms in Eq.~(\ref{eq:sell}) are of the same order in the $\ell\rightarrow\infty$ limit.

We may expand both $a(g)$ and $\tilde{S}_{0}(x)$ into polynomials near the QCP,
\begin{gather}
a(g) = \sum_{k=0}^{k_{\mr{max}}}\frac{a^{(k)}(g_{c})}{k!}(g-g_{c})^{k}, \\
\tilde{S}_{0}(x) = \sum_{l=0}^{l_{\mr{max}}}\frac{\tilde{S}_{0}^{(l)}(0)}{l!}x^{l},
\end{gather}
and take the critical point $g_{c}$, the critical exponent $\nu$, and the coefficients $a^{(k)}(g_{c})$ and $\tilde{S}_{0}^{(l)}(0)$ as fitting parameters. The coexistence of singular and nonsingular terms of the same order makes the fitting procedure often unstable in practice, making it quite difficult to capture the universal critical behavior in the EE through data collapse analysis.

\begin{figure}[!tb]
\centering
\includegraphics[width=\textwidth]{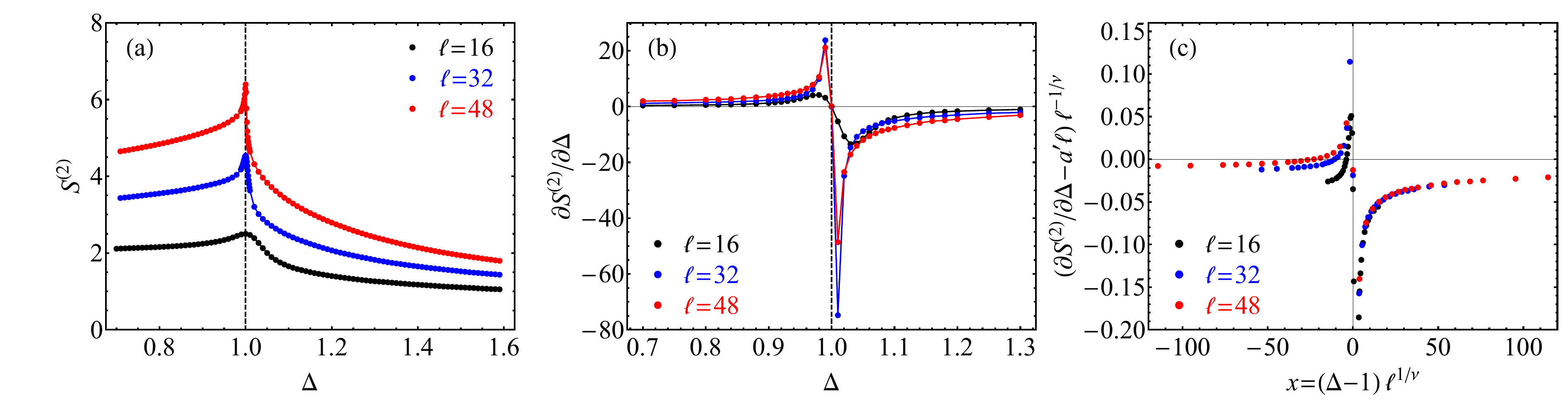}
\caption{Failed scaling analysis of DEE data near the first-order transition of the anisotropic Heisenberg model. (a) The second R\'enyi EE and (b) its derivative near the transition. (c) The DEE data cannot collapse onto a single curve with the one-parameter scaling analysis near the first-order transition.}
\label{fig:dee-first-order}
\end{figure}

\section{Failed attempt at scaling analysis near first-order transition}

The spin-1/2 anisotropic Heisenberg model on a square lattice is given by
\begin{equation}
H=J\sum_{\langle ij\rangle}\big(S_{i}^{x}S_{j}^{x}+S_{i}^{y}S_{j}^{y}+\Delta S_{i}^{z}S_{j}^{z}\big),
\end{equation}
in which the summation is taken over all n.n. bonds, and the anisotropy parameter $\Delta$ is the tuning parameter. At $\Delta=1$, the Heisenberg interaction is isotropic with an O(3) symmetry, which is spontaneously broken by the AF order at the ground state. Away from this point, the symmetry is explicitly reduced to $\mr{O}(2)=\mr{SO}(2)\rtimes \mbb{Z}_{2}$ generated by the uniaxial spin rotation and the spin-flip transformations. For $0<\Delta<1$, the ground state has an easy-plane AF order and breaks the SO(2) spin rotation symmetry, while for $\Delta>1$, the ground state has an easy-axis AF order and breaks the $\mbb{Z}_{2}$ spin flip symmetry. At $\Delta=1$, the easy-plane and the easy-axis AF order parameters coexist and jump to zero across the transition. Therefore, this point is a first-order transition with spontaneously broken O(3) symmetry~\cite{Zhao2019}.

The second R\'enyi EE and its derivative (DEE) near the first-order transition are plotted in Figs.~\ref{fig:dee-first-order}~(a) and (b). The EE shows a sharp peak at the transition point,  indicating stronger quantum correlations than the adjacent phases. This can be attributed to a larger number of Goldstone modes and a higher low-energy density of states (DOS) at the transition point. Consequently, the DEE abruptly changes sign across the transition. Following the same fitting procedure adopted in the Ising transition case, we first fit the scaling relation [Eq.~(2) in the main text] to the DEE data on the right-hand side of the transition ($\Delta>1$) but with a given $\Delta_{c}=1$, and obtain an exponent $\nu=0.53(5)$. However, we find the DEE data on the left-hand side ($\Delta<1$) cannot collapse onto a single curve with the same $\nu$ [see Fig.~\ref{fig:dee-first-order}~(c)]. Therefore, the scaling relation cannot capture the DEE data, which is not surprising because of the first-order nature of the transition.

\section{EE in Fig. 2 of the main text is rescaled by the system size $l$}

EE in Fig. 2 of the main text is rescaled by the system size $l$ are shown in  Fig.~\ref{fig:re}.

\begin{figure}[!tb]
\centering
\includegraphics[width=\textwidth]{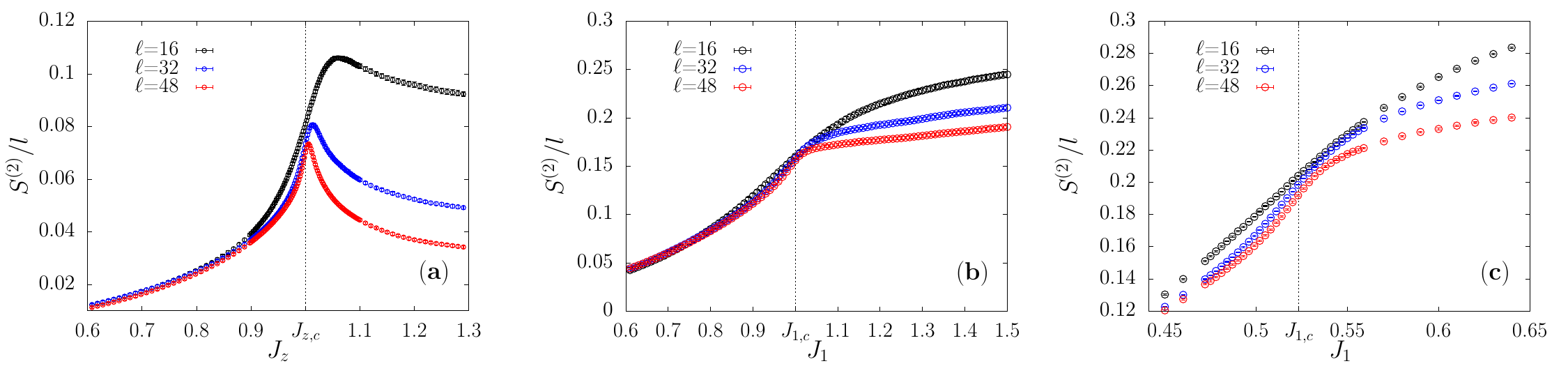}
\caption{(a) The second R\'enyi EEs rescaled by the system size $l$ near the Ising transition of the bilayer Ising-Heisenberg model. In the calculations, $J=3.045$ is fixed, while $J_{z}$ is tuned near the QCP $J_{z,c}=1$. (b) Data of the dimerized Heisenberg model with the anisotropy parameter $\Delta=0.9$. $J_{2}=2.1035$ is fixed, while $J_{1}$ is tuned around the O(2) QCP $J_{1,c}=1$. (c) Data of the dimerized isotropic Heisenberg model. $J_{2}=1$ is fixed, while $J_{1}$ is tuned around the O(3) QCP $J_{1,c}=0.52337$.}
\label{fig:re}
\end{figure}

\section{Specific parameters for DEE scaling fits in continuous phase transitions}

In the main text we find
\begin{equation}
\frac{\pd S(\ell,g)}{\pd g}= a'(g)\ell + \tilde{S}'_{0}(x)\ell^{1/\nu}.
\label{eq:dell}
\end{equation}
 where the scaling function $\tilde{S}_{0}'(x)$ can be expanded into a polynomial near the QCP and the fitted coefficients $a'$ and $\tilde{S}_{0}^{(k)}(0)$ are provided in the Table ~\ref{extm}.
\begin{equation}
\tilde{S}_{0}'(x)=\sum_{k=0}^{k=5}\frac{1}{k!}\tilde{S}_{0}^{(k)}(0)x^{k}.
\end{equation}

\begin{ruledtabular}
\begin{table}[!h]
\caption{ The fitted coefficients $a'$ and $\tilde{S}_{0}^{(k)}(0)$ near the (2+1)D Ising, O(2), and O(3) critical points correspond to Fig. 2 in the main text. When the numerical value is smaller than $10^{-5}$, we count it as zero.}
\begin{tabular}{l c c c c  c c c }
QCPs &$a'$  & $\tilde{S}_{0}^{(0)}$ 	  &$\tilde{S}_{0}^{(1)}$	&$\tilde{S}_{0}^{(2)}$ &$\tilde{S}_{0}^{(3)}$ &$\tilde{S}_{0}^{(4)}$ &$\tilde{S}_{0}^{(5)}$ \\
 	   	   	
\hline
 	   	   Ising  &-0.18(2)  & 0.249(7) 	  & 0.092(4) 	&0.019(2) &0.0022(3) &0.00013(2) &0 \\
 	   	   	
\hline
  O(2)  &0.32(3)  & 0.041(7) 	  & -0.007(3) 	&-0.007(3) &0.0017(9) &0 &0 \\

  \hline
  O(3) &0.28(9) &0.09(5)  & -0.011(8) 	  & -0.006(6) 	&0 &0 &0  \\

\end{tabular}
\label{extm}
\end{table}
\end{ruledtabular}

\end{widetext}

%\bibliography{EEScaling}
\end{document}